\documentclass[Journal,a4paper,InsideFigs, SectionNumbers, NoLineNumbers, SingleSpace]{ascelike-new}
\usepackage{xpatch}
\xapptocmd{\Section}{\addcontentsline{toc}{section}{#1}}{}{}
\usepackage[utf8]{inputenc}
\usepackage[T1]{fontenc}
\usepackage{lmodern}
\usepackage{graphicx}
\usepackage[figurename=Fig.,labelfont=bf,labelsep=period]{caption}
\usepackage{subcaption}
\usepackage{amsmath}
\usepackage{newtxtext,newtxmath}
\usepackage[colorlinks=true,citecolor=black,linkcolor=black,urlcolor=blue]{hyperref}
\usepackage{xcolor}
\begin{document}

\title{Measuring the effective stress parameter using the multiphase lattice Boltzmann method and investigating the source of its hysteresis}

\author[1,*]{Reihaneh Hosseini}
\author[1]{Krishna Kumar}

\affil[1]{Department of Civil, Architectural and Environmental Engineering, The University of Texas at Austin, USA.}

\maketitle

\renewcommand*{\thefootnote}{\fnsymbol{footnote}}
\footnotetext[1]{This author is now affiliated with Virginia Tech, Blacksburg, Virginia, USA. Email: reihos@vt.edu.}

\renewcommand*{\thefootnote}{\arabic{footnote}}
\setcounter{footnote}{0}

\begin{abstract}
The effective stress parameter, $\chi$, is essential for calculating the effective stress in unsaturated soils. Experimental measurements have captured different relationships between $\chi$ and the degree of saturation, $S_r$; however, they have not been able to justify the particular shapes of the $\chi$-$S_r$ curves. Theoretical solutions express $\chi$ as a function of $S_r$ and the air-water interfacial area, $a_{wn}$; however, $a_{wn}$ is difficult to predict, limiting further investigation of $\chi$ variation. We seek an alternative approach for studying $\chi$ by simulating the pore-scale distribution of the two fluid phases in unsaturated soils using the multiphase lattice Boltzmann method (LBM). We develop an algorithm for measuring $\chi$ based on the suction and surface tension forces applied to each grain. Using this algorithm, we simulate the $\chi$-$S_r$ curve over a full hydraulic cycle for a synthetic 3D granular soil column with immobile grains. We find that $\chi=1$ at $S_r=1$ and $\chi=0$ at $S_r=0$, while $\chi>S_r$ for all other saturations. The maximum divergence of $\chi$ from $S_r$ happens at the transition from/to the pendular regime. We also observe that the $\chi$-$S_r$ curve is hysteretic; $\chi$ is larger during wetting (imbibition) compared to drying (drainage) due to larger contribution of surface tension forces. 
\end{abstract}


\section{Introduction}
The concept of effective stress, $\sigma^\prime$, in saturated soils was first introduced by Terzaghi \citeyear{terzaghi_shearing_1936} as the stress carried by the soil skeleton, calculated as $\sigma^\prime = \sigma - u$, where $\sigma$ is the total stress and $u$ is the pore pressure. Since then, $\sigma^\prime$ has become the cornerstone of geotechnical engineering analysis of saturated soils. Later, Bishop \citeyear{bishop_principle_1959} proposed a modification to Terzaghi’s effective stress equation for unsaturated soils, $\sigma^\prime = \sigma - u_a + \chi(u_a-u_w)$, where $u_a$ and $u_w$ are pore air and pore water pressures and $\chi$ is the effective stress parameter. Bishop’s equation describes the effective stress in unsaturated soil as the summation of the net normal stress, $\sigma - u_a$, and a contribution from the matric suction, $u_a-u_w$, where the extent of the contribution depends on the material variable $\chi$. To recover Terzaghi’s original effective stress, $\chi$ should be zero for dry soil and one for fully saturated soil, but how $\chi$ should vary between zero and one is unclear.

During the early stages of developing the effective stress concept, researchers conducted experimental measurements of $\chi$ \cite{bishop_experimental_1961,jennings_limitations_1962,bishop_aspects_1963}. These measurements involved plotting $\chi$ as a function of the degree of saturation, $S_r$. Although the experimental results generally showed a decreasing trend of $\chi$ with decreasing $S_r$, no formulation was provided due to the variability in the curve’s shape. Later, Khalili and Khabbaz \citeyear{khalili_unique_1998} compiled a larger experimental data set and observed a linear trend for $\chi$ when plotted against suction (in log scale) rather than $S_r$. They proposed a formulation for $\chi$ as a function of suction and the air entry value based on the best fit to the data. They later noted that the same formulation could be used for the wetting path if the air entry value is replaced by the air expulsion value \cite{khalili_effective_2004} and also provided a new formulation for the scanning curves \cite{khalili_influence_2010}. While these formulations have been useful for developing constitutive models \cite{loret_effective_2002}, they are empirical in nature and do not offer insight into the factors influencing $\chi$.

From a theoretical perspective, it is possible to derive $\chi$ based on thermodynamic considerations. Using different thermodynamic approaches, researchers have derived $\chi= S_r$ \cite{hassanizadeh_mechanics_1990,hutter_thermodynamically_1999}. While $\chi= S_r$ is a good estimate, it does not fully align with the experimental observations. Nikooee et al. \citeyear{nikooee_effective_2013} suggested that the shortcoming of the previous thermodynamic approaches in estimating $\chi$ were due to ignoring the interfacial effects in their formulation. Following the thermodynamic approach of Hassanizadeh and Gray \citeyear{hassanizadeh_mechanics_1990}, Nikooee et al. derived $\chi$ as a function of both $S_r$ and specific air-water interfacial area, $a_{wn}$, i.e., area per unit volume. Similarly, Likos \citeyear{likos_effective_2014} revisited the $\chi$ derived by Lu et al. \citeyear{lu_closed-form_2010} and, rather than ignoring the interfacial affects as done in Lu et al., adopted the complete expression that included $a_{wn}$. These updated formulations for $\chi$ require estimating $a_{wn}$ at different saturations. Nikooee et al. used the $a_{wn}$ formulation proposed by Joekar-Niasar et al. \citeyear{joekar-niasar_insights_2008} based on pore-network modeling and found the required coefficients by curve-fitting to experimental data. Likos formulated $a_{wn}$ for an idealized unit pore with simple-cubic packing and integrated it based on the pores size distribution of the soil to find the $a_{wn}$ for the entire soil matrix, assuming that the soil consists of a collection of these idealized pores with different sizes. Culligan et al \citeyear{culligan_interfacial_2004} measured $a_{wn}$ experimentally using X-ray microtomography, but their data set was sparse and only for glass beads. Other experimental techniques for measuring $a_{wn}$, such as interfacial tracer and surfactant methods, are rather complicated to perform, and it is not always clear which interfaces have been measured \cite{costanza-robinson_air-water_2002}. Although theoretical approaches have provided greater insight into $\chi$, such as its dependence on both $S_r$ and $a_{wn}$, the difficulty in accurately determining $a_{wn}$ has limited the ability to study the variation in $\chi$ under different conditions.

Numerical modeling is an alternative to experimental or theoretical approaches for studying the effective stress and $\chi$ in unsaturated soils. Particularly of interest is a numerical modeling technique that can simulate the distribution of different fluid phases inside the pore structure of unsaturated soils and provide information on interfacial areas, among other fluid phase distribution statistics. The multiphase lattice Boltzmann method (LBM) is a good candidate for such application \cite{pan_lattice-boltzmann_2004,richefeu_lattice_2016,li_lattice_2018,hosseini_investigating_2024}. Multiphase LBM has become a popular tool for studying multiphase flow through porous media \cite{li_pore-scale_2005,ghassemi_numerical_2011,hao_pore-scale_2010,zhang_lattice_2016} as well as capillary and hydraulic behavior of unsaturated soils \cite{porter_lattice-boltzmann_2009,galindo-torres_lattice_2013,delenne_liquid_2015,li_lattice_2018}. However, the application of LBM to studying the stresses unsaturated soils remains limited. Since multiphase LBM only simulates the fluid zone, rigorous measurement of stresses inside the soil skeleton would require coupling multiphase LBM with a numerical modeling technique for simulating the grain-grain interactions, such as the discrete element method (DEM) \cite{cundall_discrete_1979}. However, for a simplified case, it is possible to measure the variation of $\chi$ due to hydraulic loading solely using multiphase LBM data, assuming the soil grains are immobile.

This study aims to gain insight into the factors influencing the shape of the $\chi$-$S_r$ curve in granular soils, using pore- and grain-scale information from multiphase LBM simulations. Additionally, we seek to evaluate whether $\chi$ exhibits hysteresis and the potential source of the hysteresis. In Section \ref{measuring stress}, we discuss an analytical method for measuring $\chi$ from a grain-scale perspective by only considering changes in the fluid phase distributions. In Section \ref{numerical method}, we introduce the numerical algorithm we developed for measuring the suction and surface tension forces, which are required for calculating $\chi$, using multiphase LBM. Finally, in Section \ref{results}, we present numerical measurements of $\chi$ for a full hydraulic cycle and discuss the mechanisms behind the trends, accompanied by comparisons with theoretical solutions. 

\section{Measuring the Effective Stress Parameter using Multiphase LBM}
\label{measuring stress}
\begin{figure}
\centering
\includegraphics[]{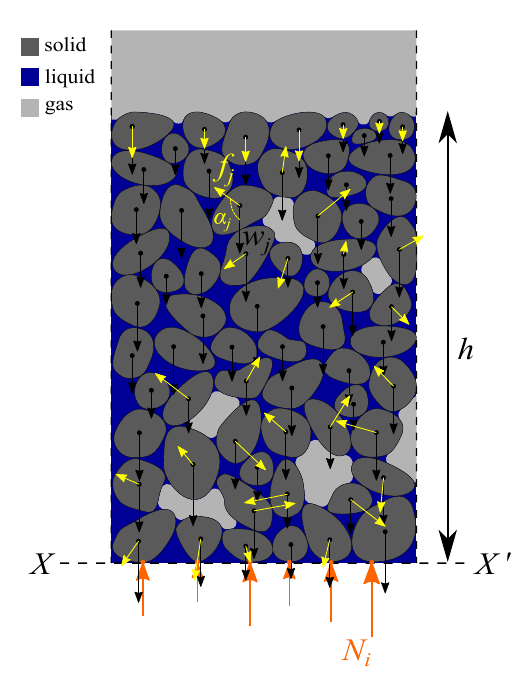}
\caption{Schematic of an unsaturated granular soil column. $w_j$ is the weight of grain $j$, and $f_j$ is the force applied by the fluid to grain $j$ due to suction and surface tension. Note that the upward buoyant forces are not shown. $N_i$ is the contact force applied to grain $i$ at the base.}
\label{fig:schematic_column}
\end{figure}

Figure~\ref{fig:schematic_column} shows a schematic of an unsaturated granular soil column. Each grain $j$ is subject to a downward force due to its weight, $w_j$, and a fluid force, $f_j$. The magnitude of $f_j$ and its direction, $\alpha_j$, depend on the suction magnitude and the distribution of the fluid phases, which vary with the degree of saturation, $S_r$. This information can be acquired from a multiphase LBM simulation. To find the vertical effective stress acting on plane $XX^\prime$ at the base of the column, the equilibrium of forces in the vertical direction can be written for the soil skeleton as

\begin{equation}
\Sigma w_j + \Sigma f_j cos(\alpha_j) - F_b -\Sigma N_i = 0
\end{equation}
where $F_b$ is the upward total buoyant force, proportional to the submerged volume of the grains. The vertical effective stress can then be found as
\begin{equation}
\sigma_v^\prime=\frac{\Sigma N_i}{A}=\frac{\Sigma w_j}{A}+\frac{\Sigma f_j cos(\alpha_j)}{A}-\frac{F_b}{A}=\frac{W_{grains}}{A}+\frac{\Sigma f_j cos(\alpha_j)}{A}-\frac{F_b}{A}.
\label{eq:micro}
\end{equation}
On the other hand, the vertical effective stress in unsaturated soils is expressed as
\begin{equation}
\sigma_v^\prime=\sigma_v - u_a +\chi(u_a-u_w).
\end{equation}
$\sigma_v$ is the total stress that already includes $u_a$, therefore, $\sigma_v-u_a$ is simply the stress due to weight of the soil column, calculated by multiplying the total unit weight of soil, $\gamma_t$, by height. As a result, $\sigma_v^\prime$ can be written as
\begin{equation}
\sigma_v^\prime=\gamma_t h +\chi(u_a-u_w)=\frac{W_{grains}+W_{water}}{A} +\chi(u_a-u_w).
\label{eq:macro}
\end{equation}
where $W_{water}$ is the weight of water within the pores. By comparing Eq.~\ref{eq:micro} and Eq.~\ref{eq:macro}, we find
\begin{equation}
\frac{W_{water}}{A} +\chi(u_a-u_w)=\frac{\Sigma f_j cos(\alpha_j)}{A}-\frac{F_b}{A}.
\end{equation}
Therefore,
\begin{equation}
\chi=\frac{\Sigma f_j cos(\alpha_j)-F_b-W_{water}}{A(u_a-u_w)}.
\label{eq:chi_general}
\end{equation}
While for a large soil column, the $F_b$ and $W_{water}$ terms in Eq.~\ref{eq:chi_general} are considerable, for the small soil sample used in this study which represents a point in the soil mass, the change of hydrostatic pressure due to the height of the specimen is negligible compared to the induced suction. Therefore, for simplicity, we do not consider gravity in the simulations, which simplifies Eq.~\ref{eq:chi_general} to 
\begin{equation}
\chi=\frac{\Sigma f_j cos(\alpha_j)}{A(u_a-u_w)}.
\label{eq:chi}
\end{equation}

This method of measuring $\chi$ has the advantage of relying only on fluid information, i.e., $f_j$, $\alpha_j$, and $u_a-u_w$, all of which can be obtained from multiphase LBM simulations. However, it is limited to measuring the vertical effective stress at the base of the soil column\footnote{Emphasis is placed on measurements at the base of the column, rather than generalizing to any horizontal plane at any depth, to avoid wavy planes. Considering the effective stress as the intergranular or contact stress, the plane for which the effective stress is found should only pass through contact points and not cut through grains. As a result, a plane that is drawn at an arbitrary depth will become wavy. At a large scale, the wavy plane will be indistinguishable from a true plane (error of 1-3\%, \citeNP{craig_craigs_2004}), but the waviness is nonnegligible at the scale of the simulations in this study. Therefore, the measurements can only be made at the base where the grain contact points are all on the same plane due to the fixed boundary condition.}. Future extensions of this work with focus on utilizing coupled LBM-DEM to determine effective stress and $\chi$ directly from contact forces \cite{younes_dem-lbm_2023}, thereby eliminating the need for Eq.~\ref{eq:chi}. In this approach, the fluid forces calculated from the multiphase LBM will be transferred to the grains, resulting in grain movements and updated contact forces. Using these updated contact forces, the contact stress tensor for a representative elementary volume (REV) can be calculated and considered as the effective stress tensor \cite{jiang_insight_2004,scholtes_micromechanics_2009,duriez_stress_2016,badetti_shear_2018,liu_micro-mechanical_2020,younes_dem-lbm_2023}, thereby generalizing the measurement of effective stress.

\section{Numerical Method}
\label{numerical method}
\subsection{Multiphase LBM}
We use the single-component multiphase LBM formulation, which corresponds to modeling the gas and liquid phases of the same fluid with phase transition. Similar to \cite{hosseini_investigating_2024}, we use lattice units (lu) (rather than physical units) for simplicity, the Carnahan-Starling equation of state (EOS) with $a=1$ lu, $b=4$ lu, $R=1$ lu and $T=0.7T_c$ to allow larger density ratios, and solid density, $\rho_s$, of $0.25$ to create a hydrophilic surface with a contact angle of $34^{\circ}$ (see Appendix~\ref{app:contact angle} for measurement of contact angle). The liquid-gas surface tension, $\gamma_{lg}$, for this parameter selection is measured as 0.0173 lu \cite{hosseini_investigating_2024}. For more details on the multiphase LBM implementation, please refer to \cite{hosseini_investigating_2024}. This section presents the developed fluid force measurement algorithm; validation of this algorithm is presented in Appendix~\ref{app:validation}.

\subsection{Algorithm for Measuring Fluid Forces}
\label{algorithm}
\begin{figure}[b]
\centering
\includegraphics[]{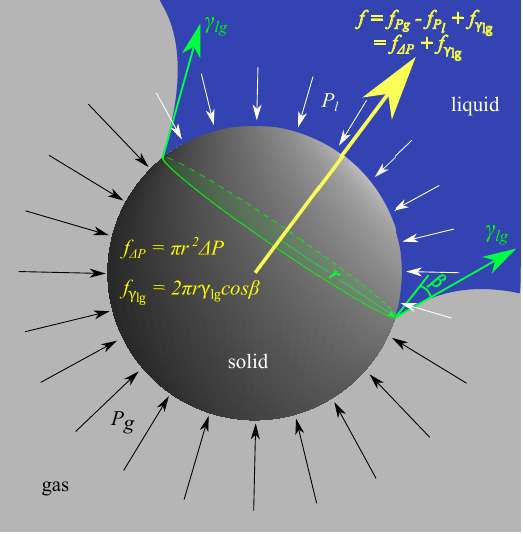}
\caption{Schematic of the fluid forces applied to a single grain.}
\label{fig:schematic_forces}
\end{figure}

\begin{figure}
\centering
\includegraphics[width = 1\textwidth]{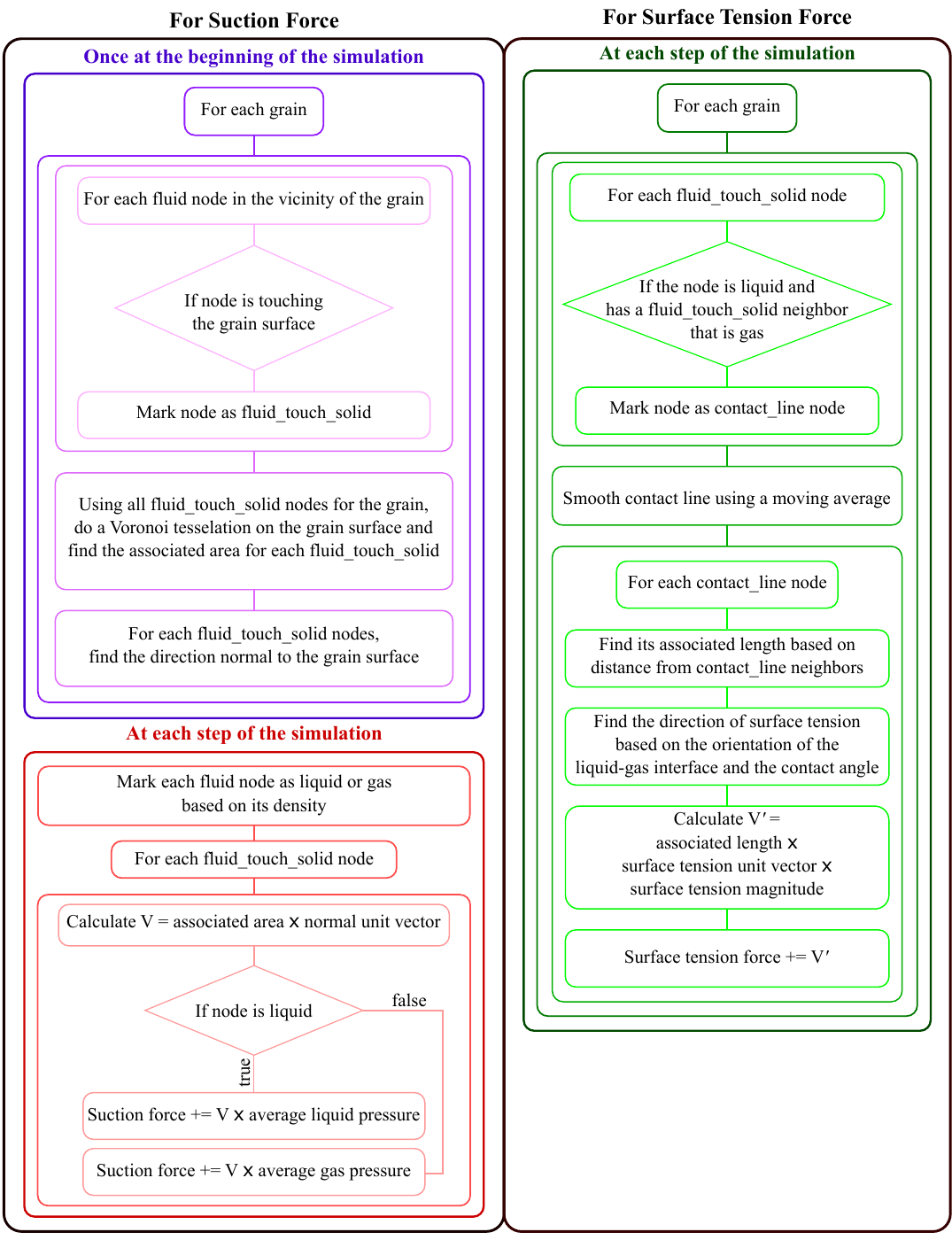}
\caption{Flowchart of the algorithm developed for measuring suction and surface tension forces in a multiphase LBM simulation.}
\label{fig:algorithm}
\end{figure}

The total fluid force, $f$, applied to a solid grain surrounded by two-phase fluid has two components, force due to suction and force due to surface tension. These components are shown schematically in Figure~\ref{fig:schematic_forces}. The suction force, $\boldsymbol{f_{\Delta P}}$, can be measured by integrating the gas and liquid pressures, $P_g$ and $P_l$ (this terminology is preferred over $u_a$ and $u_w$ for generalization), on the surface they act on. This integration is equivalent to multiplying the pressure difference, i.e., suction, $\Delta P=P_g-P_l$, by the projected area, $\pi r^2$, shown in green in Figure~\ref{fig:schematic_forces}. Similarly, the surface tension force, $\boldsymbol{f_{\gamma_{lg}}}$, can be measured by integrating the liquid-gas surface tension, $\gamma_{lg}$, over the length it acts on. This integration is equivalent to multiplying $\gamma_{lg}$ by the length of the contact line, $2\pi r$, and $cos\beta$, where $\beta$ is the angle between the liquid-gas interface and the direction normal to the projected area. These forces need to be measured during a multiphase LBM simulation to calculate $\chi$ using Eq.~\ref{eq:chi}. For this purpose, we have developed the algorithm summarized in Figure~\ref{fig:algorithm} and discussed in detail in the two following sections.

\subsubsection{Measuring Suction Forces}
In our simulations, the magnitude of the gas and liquid pressures are easily calculated by substituting the fluid densities into the EOS. However, further effort is required to find the positions where the gas and liquid pressures act and the orientation of the pressure forces. We implement the following algorithm.

\begin{figure}
\centering
\includegraphics[]{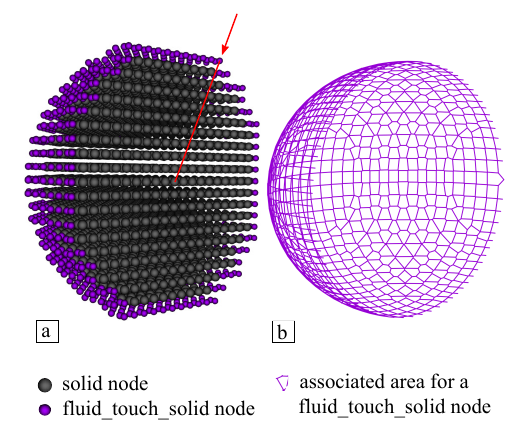}
\caption{Information required for calculating the suction force, shown for only half a spherical grain for ease of visualization. a) Fluid nodes on the grain surface (i.e., fluid\textunderscore touch\textunderscore solid nodes). The direction normal to the grain surface is shown in red for one of the fluid\textunderscore touch\textunderscore solid nodes. b) Voronoi tessellation of the grain surface using the mapped fluid\textunderscore touch\textunderscore solid nodes as seeds.}
\label{fig:suction_force}
\end{figure}

Once during model initialization:
\begin{itemize}
    \item For each grain, find the fluid nodes that are in contact with its surface. This is done by looping through all fluid nodes that surround each grain, checking their neighboring nodes, and marking them as fluid\textunderscore touch\textunderscore solid if they have a solid node neighbor. These nodes are shown in purple for half a spherical grain in Figure~\ref{fig:suction_force}a.
    \item For each fluid\textunderscore touch\textunderscore solid node, find its associated area on the surface of the grain. The associated area is needed for integrating the pressure over an area. This is done by mapping the fluid\textunderscore touch\textunderscore solid nodes onto the true surface of the grain (the surface not affected by discretization) and performing a Voronoi tessellation on it, with the mapped fluid\textunderscore touch\textunderscore solid nodes being the seeds. The tessellation for half a spherical grain is shown in Figure~\ref{fig:suction_force}b.
    \item For each fluid\textunderscore touch\textunderscore solid node, find the direction normal to the surface of the grain. The direction for one of the fluid\textunderscore touch\textunderscore solid nodes is shown in Figure~\ref{fig:suction_force}a as an example.
\end{itemize}

At each simulation step:
\begin{itemize}
    \item Loop through all fluid nodes and flag them as either liquid or gas based on a density threshold. We use a density threshold of 0.2 which is halfway between the liquid and gas densities.
    \item At each liquid/gas node that is also a fluid\textunderscore touch\textunderscore solid node, multiply its associated area by the average liquid/gas pressure and the unit vector normal to the surface to get the pressure force vector at that node.
    \item Sum the pressure force vectors at all fluid\textunderscore touch\textunderscore solid nodes for a particular grain to get the suction force applied from the fluid to that grain, $\boldsymbol{f_{\Delta P}}$.
\end{itemize}

\subsubsection{Measuring Surface Tension Forces}
\label{surface tension force}
Similar to calculating the suction force, for calculating surface tension force, the positions where the surface tension acts and its orientation at each position need to be identified. We implement the following algorithm.

\begin{figure}[t]
\centering
\includegraphics[]{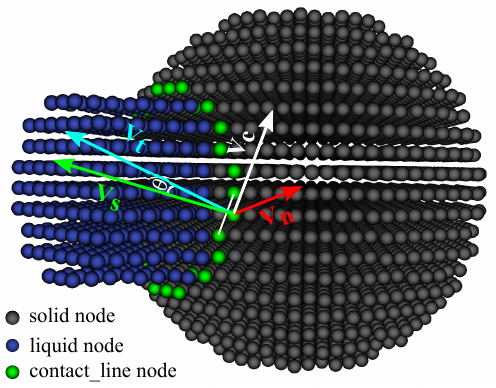}
\caption{Finding the orientation of the surface tension vector for a node on the contact line. The contact line is shown in green, while the liquid bridge is shown in blue. $\boldsymbol{V_n}$ is the vector normal to the surface of the grain at the selected node. $\boldsymbol{V_c}$ is the vector connecting the two contact line neighbors of nodes, scaled and shifted for better visualization. $\boldsymbol{V_t}$ is the result of the cross product of $\boldsymbol{V_n}$ and $\boldsymbol{V_c}$; therefore, it is tangent to the surface of the grain and perpendicular to the contact line. $\boldsymbol{V_s}$ is the correct orientation of the surface tension resulting from rotating $\boldsymbol{V_t}$ by an angle $\theta$ away from the grain surface while remaining perpendicular to the contact line.}
\label{fig:surface_tension_force}
\end{figure}

At each step of the simulation:
\begin{itemize}
\item For each grain, find the nodes on the contact line between liquid, gas, and solid phases. This is done by looping through all fluid\textunderscore touch\textunderscore solid nodes that have been flagged as liquid, checking their neighbors, and marking them as contact\textunderscore line nodes if they have a gas neighbor. Also, a smoothing function is applied to the identified contact\textunderscore line nodes to reduce the jaggedness due to discretization. See the green points in Figure~\ref{fig:surface_tension_force}.
\item For each contact\textunderscore line node, find its associated length. This is done by searching through the neighbors of each contact\textunderscore line node, finding the two neighbors that are also on the contact line, and taking half the distance between the two neighbors as the associated length for that node. See the white line in Figure~\ref{fig:surface_tension_force}.
\item For each contact\textunderscore line node, find the direction of the surface tension. To this end, first the vector connecting the two contact\textunderscore line neighbors,$\boldsymbol{V_c}$, is calculated, corresponding to the vector tangent to the contact line at the given node. Then, the cross product of $\boldsymbol{V_c}$ and the vector normal to the surface of the grain, $\boldsymbol{V_n}$, is calculated, which corresponds to a vector perpendicular to the contact line, and tangent to the surface of the grain. Let us call this new vector the tangent vector, $\boldsymbol{V_t}$, referring to it being tangent to the surface of the grain. Then, the state of a node in the direction of the tangent vector is checked; if the node is liquid, then the direction is correct; if the node is gas, then the direction is flipped 180$^{\circ}$. If the contact angle is zero, this tangent vector corresponds to the correct direction of the surface tension applied to the grain. When the contact angle is not zero, this vector needs to be rotated away from the grain inside a plane perpendicular to the contact line, to correspond to the correct orientation of the surface tension vector, $\boldsymbol{V_s}$.
\item At each contact\textunderscore line node, multiply its associated length by the magnitude of the surface tension and the surface tension unit vector, $\boldsymbol{V_s}/|\boldsymbol{V_s}|$, to get the surface tension force vector at that node.
\item Sum the surface tension force vectors at all contact\textunderscore line nodes for a particular grain to get the surface tension force applied from the fluid to that grain, $\boldsymbol{f_{\gamma_{lg}}}$.
\end{itemize}

\section{Results}
\label{results}
\subsection{Model Configuration}
\begin{figure}[h!]
\centering
\includegraphics[width = 0.4\textwidth]{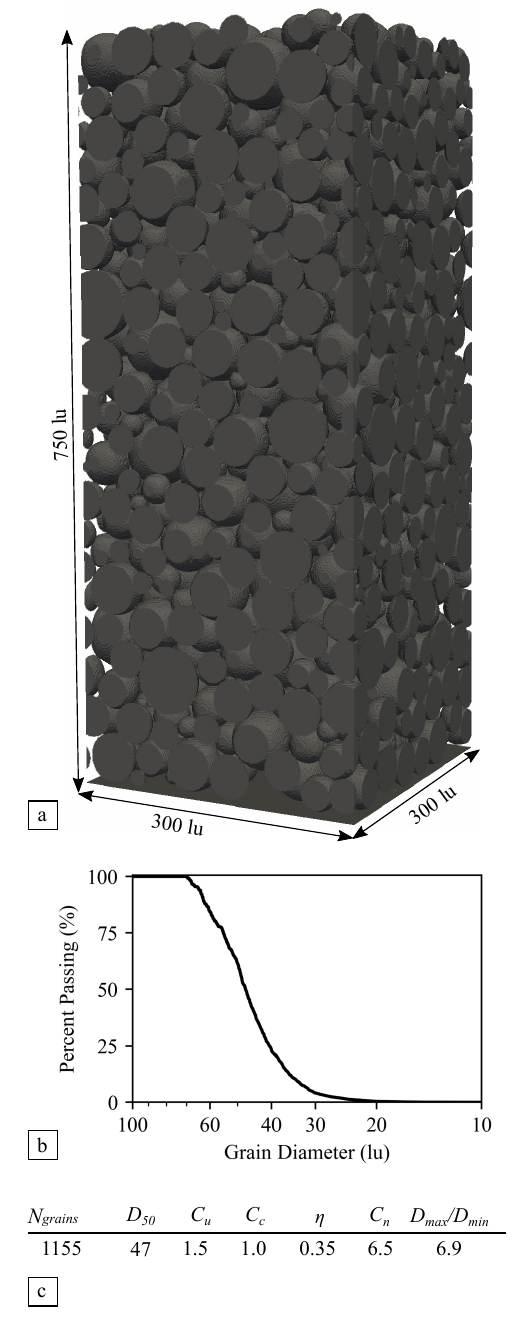}
\caption{3D granular soil column model. a) Visualization of the granular packing. b) Grain
size distribution curve. c) Statistics of the granular packing including total number of grains $N_{grain}$, diameter corresponding to 50\% passing of the grain size distribution $D_{50}$,  coefficient of uniformity $C_u = D_{60}/D_{10}$, coefficient of curvature $C_c = D_{30}^2/(D_{60}D_{10})$, porosity $\eta$ = pore volume / total volume, coordination number $C_n = 2\times$number of contacts / $N_{grains}$, and polydispersity = $D_{max}/D_{min}$.}
\label{fig:model_config}
\end{figure}

We use the 3D granular soil column model consisting of spherical grains shown in Figure~\ref{fig:model_config}a to calculate the effective stress parameter using Eq.~\ref{eq:chi}. We use the PFC3D software, based on DEM, to create a stable granular packing. We generate grains following the grain size distribution in Figure~\ref{fig:model_config}b within a fixed-size domain to match the target porosity \cite{hosseini_investigating_2024}. We then run the DEM simulation until the packing is stable. The boundary conditions are periodic in the horizontal directions, closed with a solid plate at the bottom and open at the top. Figure~\ref{fig:model_config}c summarizes other properties of the granular packing.   
 
\subsection{Hydraulic Path and SWCC}
\begin{figure}
\centering
\includegraphics[width=1\textwidth]{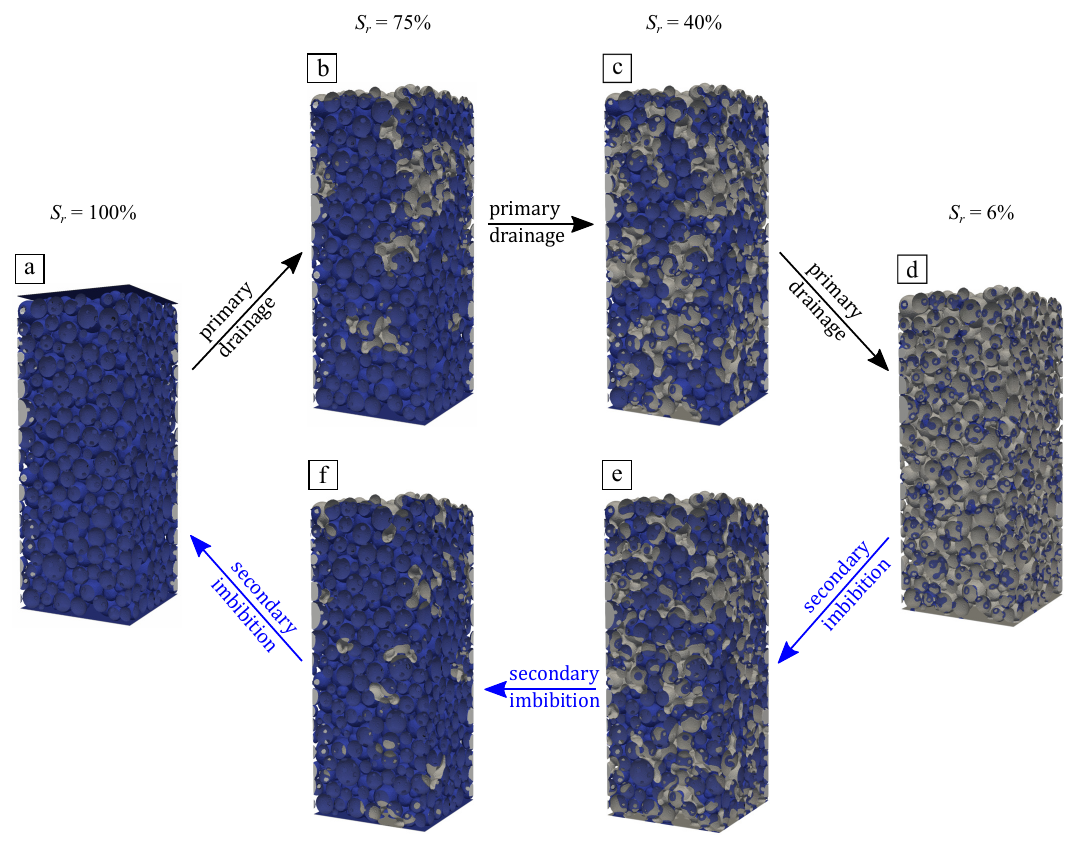}
\caption{Hydraulic path simulated for the 3D granular soil column model.}
\label{fig:hydraulic_path}
\end{figure}

\begin{figure}
\centering
\includegraphics[]{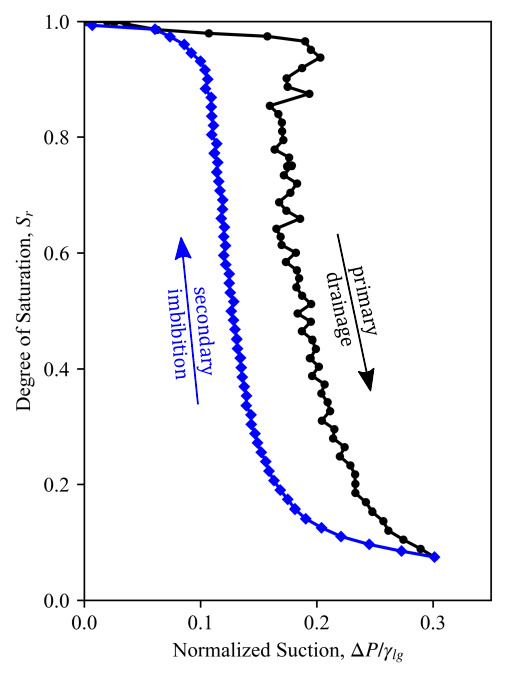}
\caption{Simulated soil-water characteristics curve (SWCC) for the 3D granular soil column model.}
\label{fig:swcc}
\end{figure}

We initialize the multiphase LBM simulation from full saturation of the soil column, shown in Figure~\ref{fig:hydraulic_path}a. We then uniformly drain the liquid to a $S_r$ of 6\%, Figure Figure~\ref{fig:hydraulic_path}a to Figure~\ref{fig:hydraulic_path}d, to get the primary drainage curve. Subsequently we inject liquid back to full saturation, Figure~\ref{fig:hydraulic_path}d to Figure~\ref{fig:hydraulic_path}a, to get the secondary imbibition curve. We also simulate scanning curves starting from intermediate saturations. Figure Figure~\ref{fig:swcc} shows these curves that together form the soil water characteristic curve (SWCC) for this soil column. 

The drainage and injection of liquid is done similar to \citeN{hosseini_investigating_2024} by changing the liquid density everywhere in the system by a small amount and allowing it to reach equilibrium under the new conditions. This procedure corresponds to a uniform change of $S_r$ in the specimen and is physically analogous to placing a specimen in a humidity chamber and decreasing/increasing the humidity. It is also possible to drain/inject from the bottom of the soil column, which is more common in physical experiments; however, changing $S_r$ from the boundary creates a gradient of suction and $S_r$ along the height of the column and introduces undesired variability. Therefore, we use the uniform-$S_r$-change procedure in this study.

\subsection{Suction and Surface Tension Forces as a Function of the Degree of Saturation}
\begin{figure}
\centering
\includegraphics[width=\textwidth]{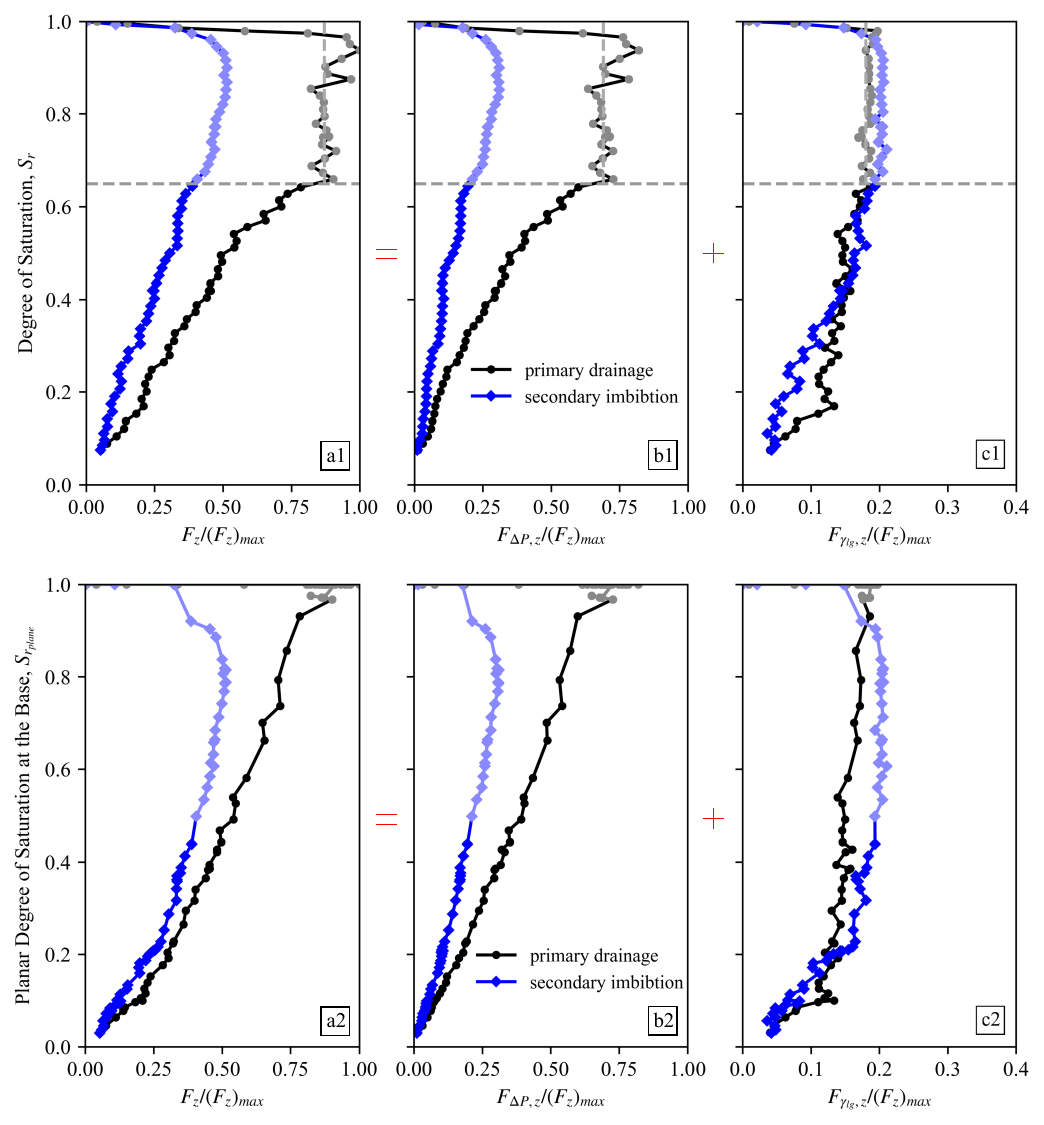}
\caption{The measured sum of vertical fluid forces, $F_z$, as well as individual contributions of suction forces, $F_{\Delta P,z}$, and surface tension forces, $F_{\gamma_{lg},z}$, as a function of 1) the global degree of saturation, $S_r$ and 2) the degree of saturation of the plane at the base of the soil column, $S_{r_{plane}}$, for the 3D granular soil column model. Points with $S_r>0.65$ have been marked with a different color to show how they translate to $S_{r_{plane}}$.}
\label{fig:forces}
\end{figure}

The sum of forces, $\Sigma f_j cos(\alpha_j)$, in Eq.~\ref{eq:chi}, is calculated at each step of the hydraulic path in Figure~\ref{fig:swcc}, using the sum of the vertical components of the suction and surface tension forces:
\begin{equation}
\Sigma f_j cos(\alpha_j)=F_z=\Sigma f_{\Delta P,z}+\Sigma f_{\gamma_{lg},z}=F_{\Delta P,z}+F_{\gamma_{lg},z}
\label{eq:sum_forces}
\end{equation}
These forces are shown in Figure~\ref{fig:forces} row 1. During primary drainage, the forces are initially zero at $S_r=1$ and increase rapidly until the air-entry value is reached at $S_r=0.97$. The forces then oscillate about a constant value for $0.97>S_r>0.65$, and decrease for $S_r<0.65$. A closer inspection reveals that $S_r=0.65$ is the saturation at which the gas phase reaches the bottom of the column during primary drainage. If we replot the forces as a function of the degree of saturation of the plane at the base of the soil column, $S_{r_{plane}}$, in Figure~\ref{fig:forces} row 2, we see a one-to-one relationship between the forces and $S_{r_{plane}}$, for the entire range of $S_{r_{plane}}$ during primary drainage and $S_{r_{plane}}<0.8$ during secondary imbibition. Since the summation of fluid forces in Eq.~\ref{eq:sum_forces} corresponds to the imbalanced fluid forces at the base of the column which need to be balanced by the contact forces at the base (see Section~\ref{measuring stress}), it is reasonable for these values to be a function of the $S_r$ at the base. Alternatively, if we were measuring the average contact force inside the column, it would have correlated with the global $S_r$. Therefore, we use $S_{r_{plane}}$ rather than the global $S_r$ hereafter.

Based on Eqs.~\ref{eq:chi} and \ref{eq:sum_forces} , $F_z$ divided by the cross-sectional area of the column, $A$, corresponds to $\chi(u_a-u_w)=\chi\Delta P$ which is commonly known as suction stress, although it includes contributions from both suction and surface tension. Since $A$ is constant, the trend of $F_z$ corresponds to the trend of the suction stress, therefore, based on Figure~\ref{fig:forces}a, the suction stress is hysteretic, with the drainage (drying) suction stress being higher than the imbibition (wetting). This result agrees with the predictions made by \citeN{likos_effective_2014} for sandy soil. Suction stress is a function of both $\chi$ and $\Delta P$. The hysteresis of $\Delta P$ is evident in Figure~\ref{fig:swcc}, and its source has been discussed in \citeN{hosseini_investigating_2024}. Here, we show that $\chi$ also contributes to the hysteresis of suction stress, and we investigate the factors affecting the shape of the $\chi$-$S_r$ curve and its hysteresis.

\subsection{$\chi$ as a function of the degree of saturation}

\begin{figure}
\centering
\includegraphics[width=\textwidth]{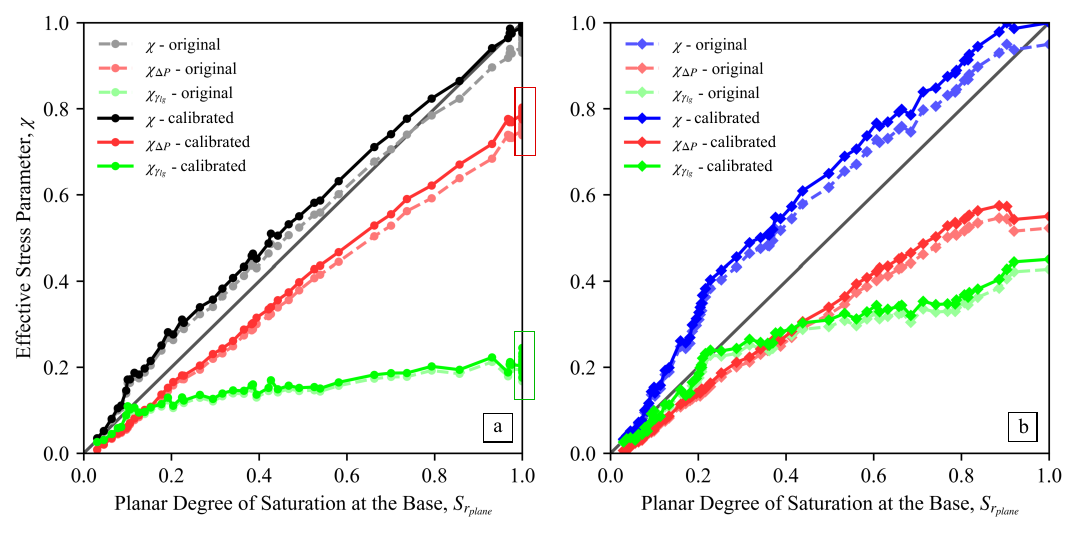}
\caption{Measured $\chi$ a function of $S_{r_{plane}}$ for the 3D granular soil column model during a) primary drainage and b) secondary imbibition.}
\label{fig:chi}
\end{figure}

A shown previously in Eq.~\ref{eq:chi}, $\chi$ corresponds to the ratio of suction stress, $F_z/A$, to suction, $\Delta P$. Figure~\ref{fig:chi} shows the $\chi$ measured in the simulation, including individual contributions of suction and surface tension forces to $\chi$:

\begin{equation}
\chi=\frac{F_z}{A\Delta P}=\frac{F_{\Delta P,z}}{A \Delta P}+\frac{F_{\gamma_{lg},z}}{A \Delta P}=\chi_{\Delta P}+\chi_{\gamma_{lg}}
\label{eq:chi_sum}
\end{equation}

Since $F_{\Delta P,z}$ already includes $\Delta P$, when divided by $A\Delta P$, the $\Delta P$ cancels out, making $\chi_{\Delta P}$ simply the ratio of the projected area to the cross-sectional area; for instance, for the single grain shown in Figure~\ref{fig:schematic_forces}, $\chi_{\Delta P}=f_{\Delta P}/(A\Delta P)=\pi r^2/A$, where $\pi r^2$ is what we refer to as the projected area and $A$ is the cross-sectional area on which the the stress is defined. Another example is shown in Figure~\ref{fig:2D_drain} for a schematic 2D representation of multiple grains (this figure is discussed in detail in Section 4.4.2.1). In this case, we are only interested in the vertical forces, therefore, only the horizontal components of the projected areas ($a_i$) are considered. Once again, we see that $\chi_{\Delta P}$ is simply $A_{projected}/A$. 

Based on the validation test results in Appendix~\ref{app:validation}, we know that in our simulations both $F_{\gamma_{lg},z}$ and the projected area are underestimated by 5\%. If we now correct all the curves in Figure~\ref{fig:chi} by a factor of 1.05, we see that the calibrated $\chi$ curves take the expected value of 1 for an $S_r$ of 1. We also observed in the validation test that the suction was overestimated, but that only applies to low saturations where the grid resolution is insufficient for the small amount of liquid content. Therefore, we expect $\chi_{\gamma_{lg}}$ to be underestimated at low $S_r$, particularly $S_r<0.2$, but we have not applied further correction here. Below, we first discuss the observed behavior for $\chi_{\Delta P}$, $\chi_{\gamma_{lg}}$ and $\chi$ during primary drainage and secondary imbibition, and subsequently, justify the observed behavior based on the differences in phase distributions.

\subsubsection{Observed Behavior of $\chi$}

\begin{figure}
\centering
\includegraphics[width=\textwidth]{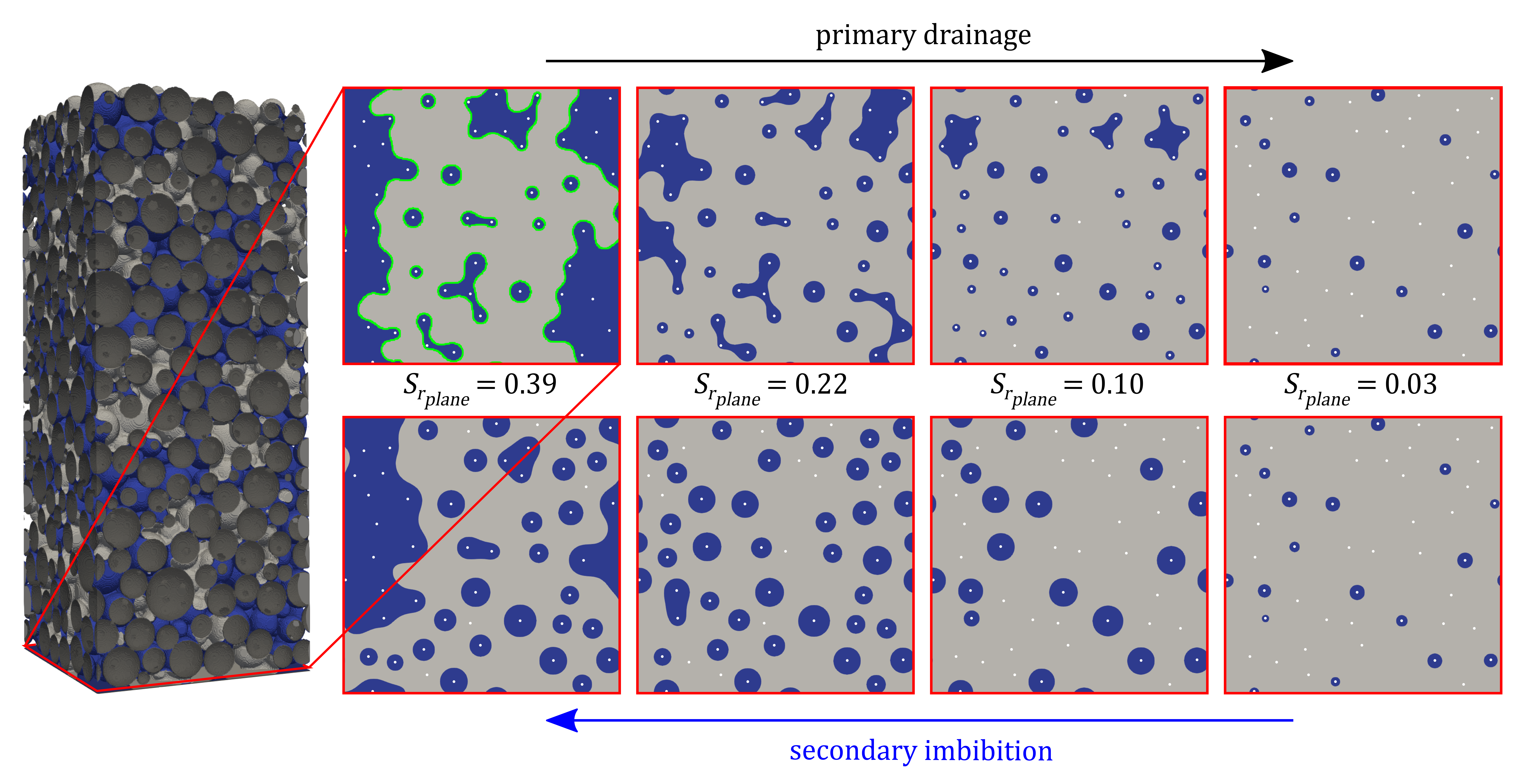}
\caption{Phase distributions on the plane at the base of the 3D granular soil column model. Blue is liquid, and gray is gas. The white dots are contact points between the grains and the base plate. The green line in the first subplot is an example of the interfacial line discussed in Section~\ref{comparison}.}
\label{fig:bottom_plane}
\end{figure}

During primary drainage, while $S_{r_{plane}}=1$ (i.e., the gas phase has not reached the bottom), $\chi_{\Delta P}$ remains at a constant value of 0.8 with minor oscillations; see the red box in Figure~\ref{fig:chi}a. Similarly, $\chi_{\gamma_{lg}}$ oscillates about a mean value of 0.2 before the gas phase reaches the bottom; see the green box in Figure~\ref{fig:chi}a. The combination of $\chi_{\Delta P}$ and $\chi_{\gamma_{lg}}$ in this range results in $\chi=1$ for $S_{r_{plane}}=1$, confirming that the grain-scale perspective using forces calculated from multiphase LBM is consistent with the continuum perspective. Once the gas phase reaches the bottom, $\chi_{\Delta P}$ decreases linearly with $S_{r_{plane}}$. Similarly, $\chi_{\gamma_{lg}}$ decreases with $S_{r_{plane}}$, but at a lower rate for $S_{r_{plane}}>0.1$ and a higher rate for $S_{r_{plane}}<0.1$, compared to $\chi_{\Delta P}$. Visualizing the simulation results indicates that the slope change of the $\chi_{\gamma_{lg}}$ curve happens when the bottom plane enters the pendular regime, i.e., liquid only in the form of bridges between grains. This is shown in Figure~\ref{fig:bottom_plane}, where the phase distributions on the plane at the base of the soil column are visualized for different saturation levels. The circular liquid regions are cross-sections of liquid bridges between the grains at the bottom of the soil column and the base plate. The other irregular-shaped liquid regions are cross-sections of liquid clusters connecting multiple grains to the base plate. At $S_{r_{plane}}=0.1$ during drainage, all but three liquid clusters are in the form of bridges. A slight reduction of $S_{r_{plane}}$ at this point results in the base plane entering the pendular regime, which coincides with the change of slope of the $\chi_{\gamma_{lg}}$ curve. As a result of the combination of $\chi_{\Delta P}$ and $\chi_{\gamma_{lg}}$ during primary drainage, $\chi$ deviates from $S_{r_{plane}}$ as $S_{r_{plane}}$ decreases from 1.0, reaches a maximum deviation of 0.07 when the base plane enters the pendular regime at $S_{r_{plane}}\simeq0.1$, and gradually converges back to $S_{r_{plane}}$ at $S_{r_{plane}}<0.1$.

During secondary imbibition, Figure~\ref{fig:chi}b, the trends of $\chi_{\Delta P}$ and $\chi_{\gamma_{lg}}$ curves are similar to primary drainage, except at $S_{r_{plane}}>0.8$ where the trends become nonlinear. Visualization of the simulation results indicates that the nonlinearity at $S_{r_{plane}}>0.8$ happens during the filling of the last pore, the reason for which will be discussed in the next section. Although the $\chi_{\Delta P}$ and $\chi_{\gamma_{lg}}$ imbibition curves show similar trends to drainage curves, their magnitudes are different. Comparing Figure~\ref{fig:chi}b to Figure~\ref{fig:chi}a, we see that $\chi_{\Delta P}$ is smaller at any given $S_{r_{plane}}$ during imbibition compared to drainage, while $\chi_{\gamma_{lg}}$ is much larger during imbibition. This is due to the particular phase distribution during imbibition, characterized by the presence of many gas clusters in the system, and will be discussed further in the next section. In addition, the slope change of $\chi_{\gamma_{lg}}$ happens at a higher $S_{r_{plane}}$ of 0.2 during imbibition, which can be again linked to the pendular regime. Looking at Figure~\ref{fig:bottom_plane}, we see that the base plane remains in the pendular regime for a wider $S_{r_{plane}}$ range during imbibition. For instance, at $S_{r_{plane}}=0.22$ during imbibition, all but one liquid cluster are in the form of large bridges (blue circles), showing that the transition from the pendular regime has just occurred. In contrast, at the same $S_{r_{plane}}$ during drainage, small bridges and liquid clusters connecting multiple grains coexist, and the regime is far from pendular. As a result of the combination of $\chi_{\Delta P}$ and $\chi_{\gamma_{lg}}$ during secondary imbibition, $\chi$ deviates from $S_{r_{plane}}$ as saturation increases, reaches a maximum deviation of 0.18 at $S_{r_{plane}}= 0.2$  where the base plane exits the pendular regimes, and gradually converges back to $S_{r_{plane}}$, reaching a value of 1 at $S_{r_{plane}}=1$.
     
\begin{figure}
\centering
\includegraphics[]{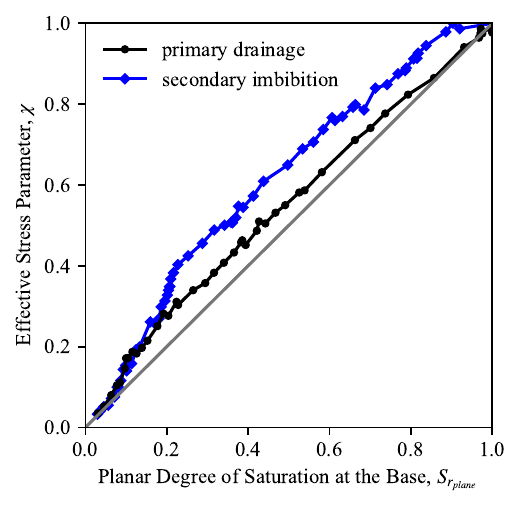}
\caption{Comparison of the drainage and imbibition $\chi$-$S_{r_{plane}}$ curves, for the 3D granular soil column model.}
\label{fig:chi_hysteresis}
\end{figure}

Figure~\ref{fig:chi_hysteresis} shows the overall $\chi$ curves for drainage and imbibition plotted together. We see that $\chi$ is hysteretic at $S_{r_{plane}}>0.1$, taking a larger magnitude during imbibition due to the larger contribution of $\chi_{\gamma_{lg}}$. At $S_{r_{plane}}<0.1$, both imbibition and drainage are in a pendular regime and have similar phase distributions; therefore, there is no hysteresis. 

\subsubsection{Conceptual Models for Understanding the Behavior of $\chi$}

\begin{figure}
\centering
\includegraphics[width=\textwidth]{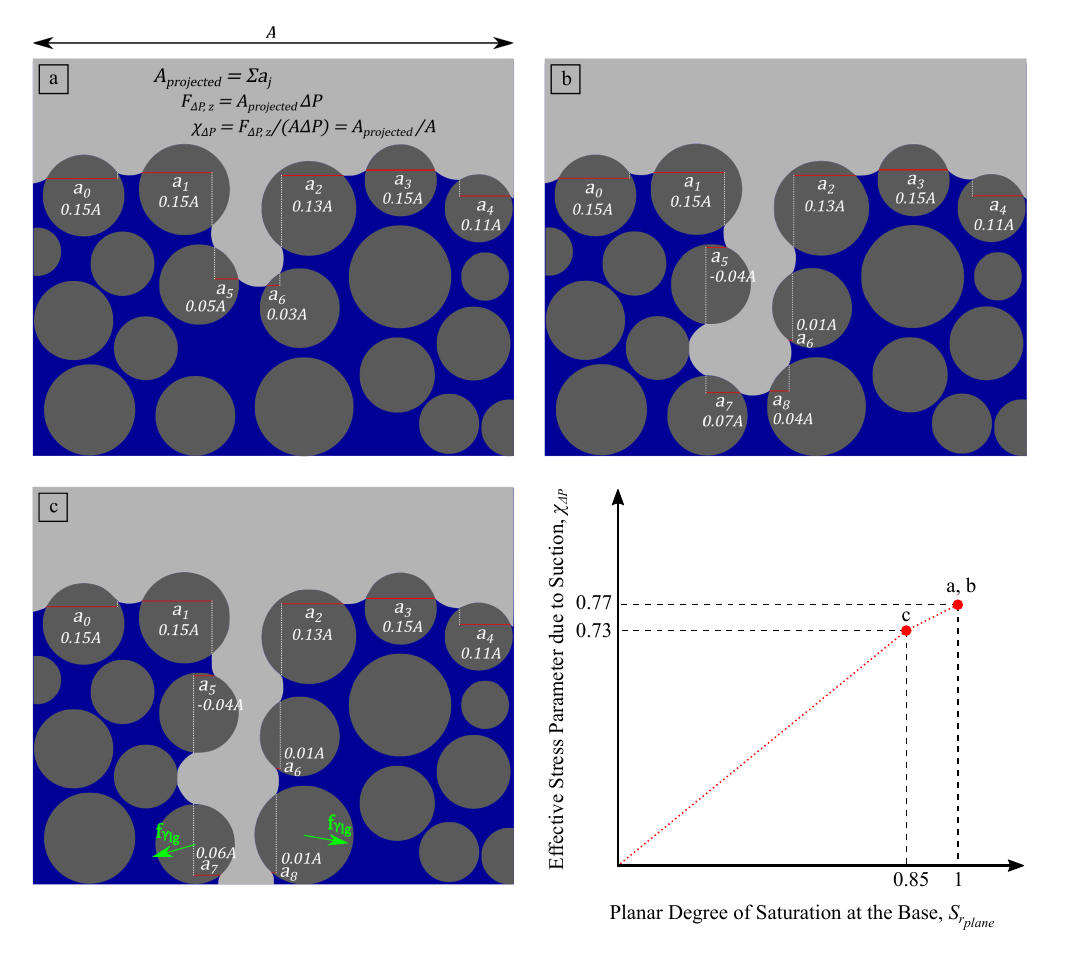}
\caption{A schematic diagram for understanding the change of $\chi_{\Delta P}$ with $S_{r_{plane}}$ during primary drainage.}
\label{fig:2D_drain}
\end{figure}

\begin{figure}
\centering
\includegraphics[width=\textwidth]{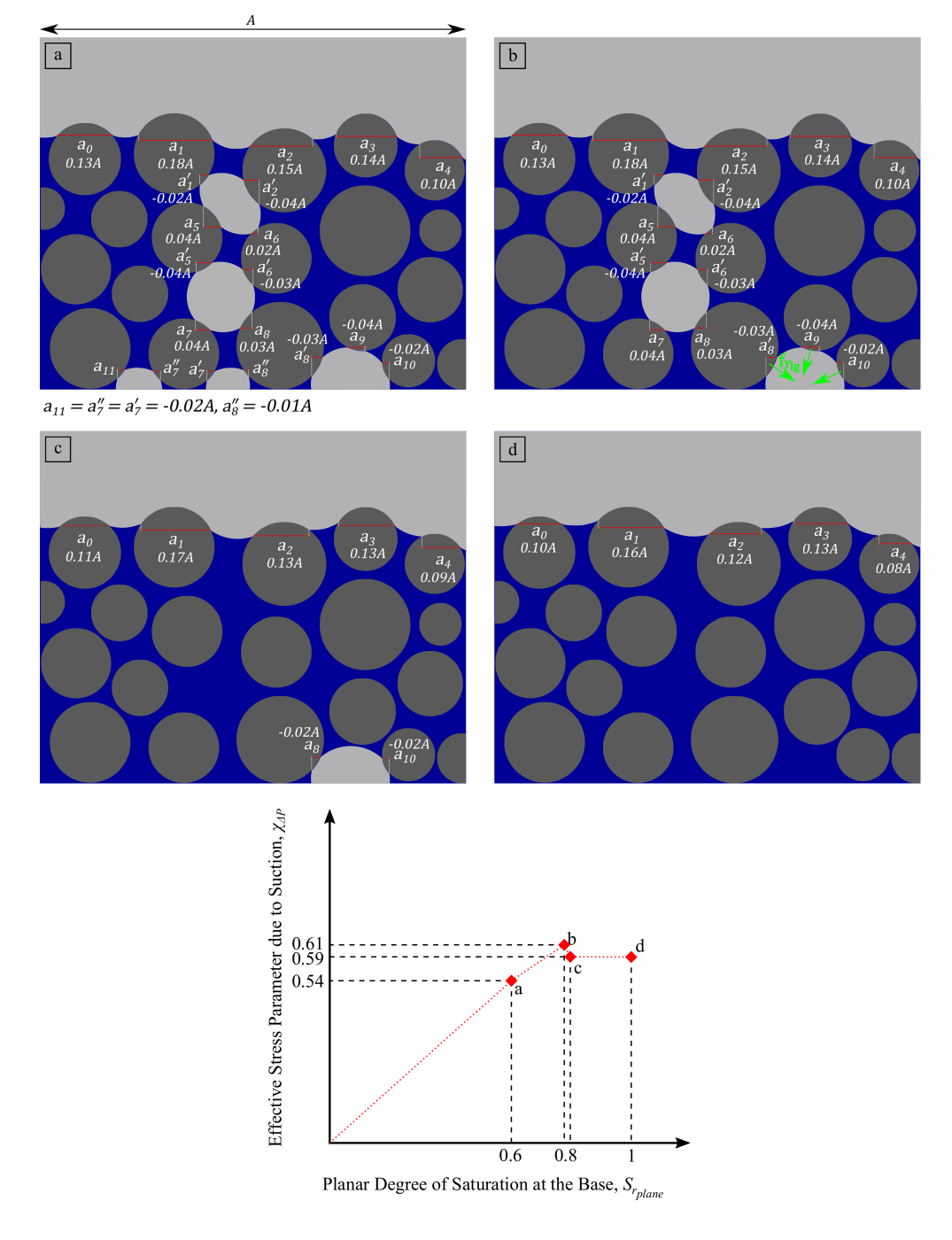}
\caption{A schematic diagram for understanding the change of $\chi_{\Delta P}$ with $S_{r_{plane}}$ during secondary imbibition.}
\label{fig:2D_imbib}
\end{figure}

In this section, we justify the observed behaviors for $\chi_{\Delta P}$ and $\chi_{\gamma_{lg}}$ using conceptual models. For $\chi_{\Delta P}$, we use the model in Figures~\ref{fig:2D_drain} and \ref{fig:2D_imbib}, which is a schematic representation of the simulated granular soil column in Figure~\ref{fig:hydraulic_path}. In this conceptual model, we have reduced the dimension of the problem from 3D to 2D for ease of visualization, included fewer grains for simplicity, and decreased the height-to-width ratio of the column to accelerate the gas phase reaching the bottom. Note that when switching from 3D to 2D, the grains need to be spaced out to allow fluid movement; this does not affect the observed behavior in this study since the grains are immobile. For $\chi_{\gamma_{lg}}$, in addition to the previous model, we use the model in Figure~\ref{fig:single_grain}, which is a schematic representation of a single grain at the bottom of the 3D soil column. These conceptual models are suitable tools for explaining micromechanisms without the need for complicated visualizations of the large 3D soil column. It is important to note that the phase distributions (i.e., the distribution of liquid and gas phases) in these schematics have been drawn based on an understanding of the pore emptying and pore filling processes during drainage and imbibition \cite{hosseini_investigating_2024}, and are consistent with the behavior observed for the simulated soil column in 3D.

\noindent{4.4.2.1   $\chi_{\Delta P}$}

As we saw previously, during primary drainage, $\chi_{\Delta P}$ remains at a constant value while $S_{r_{plane}}=1$ and subsequently decreases as $S_{r_{plane}}$ decreases. We explain this behavior using the model in Figure~\ref{fig:2D_drain}, considering that $\chi_{\Delta P}$ corresponds to the ratio of the total projected area, $A_{projected}$, to the cross-sectional area, $A$. Note that since this is a 2D schematic, areas are represented as lengths.  The schematic demonstrates that when the system's saturation changes without the gas reaching the bottom, Figures~\ref{fig:2D_drain}a and \ref{fig:2D_drain}b, the projected area of each grain, $\text{a}_\text{j}$, might change, but the sum of the individual projected areas, $\Sigma \text{a}_\text{j}=A_{projected}$, remains constant. For instance, when the gas enters a new pore in Figure~\ref{fig:2D_drain}b, $\text{a}_5$ becomes negative because the gas is pushing from underneath the grain, and $\text{a}_6$ decrease to $0.01A$; however, $\text{a}_7$ and $\text{a}_8$ are added in such a way that $\Sigma \text{a}_\text{j}$, and therefore $A_{projected}$ and $\chi_{\Delta P}$, remain constant. Once the gas phase reaches the bottom in Figure~\ref{fig:2D_drain}c, unlike in \ref{fig:2D_drain}b where $\text{a}_7$ and $\text{a}_8$ made up for the lost projected area, no new areas are added, and therefore $A_{projected}$ decreases.

During secondary imbibition, we observed that $\chi_{\Delta P}$ increases linearly with $S_{r_{plane}}$ for $S_{r_{plane}}<0.8$, while the change rate of $\chi_{\Delta P}$ decreases and even switches sign for $S_{r_{plane}}>0.8$. We explain this behavior using the conceptual model in Figure~\ref{fig:2D_imbib}, which is the same as in Figure~\ref{fig:2D_drain}, but during imbibition. Based on the understanding of the phase distribution from \citeN{hosseini_investigating_2024}, we know that during imbibition, there are many gas clusters inside the soil matrix, unlike during drainage, where there is only a single gas cluster connected to the top. As seen in Figure~\ref{fig:2D_imbib}a, the projected areas due to the gas clusters in the body of the soil effectively cancel out. Therefore, the controlling projected areas are at the top and bottom. At low saturations, i.e., Figures~\ref{fig:2D_imbib}a and \ref{fig:2D_imbib}b in the conceptual model and $S_{r_{plane}}<0.8$ in the actual model, the effect of the projected areas at the bottom is more pronounced. As $S_{r_{plane}}$ increases, from \ref{fig:2D_imbib}a to \ref{fig:2D_imbib}b, the pores at the bottom start filling. This results in the negative projected areas (i.e., upward push from the gas phase) disappearing, thus, $A_{projected}$ and $\chi_{\Delta P}$ increasing. At higher saturations, i.e., Figures \ref{fig:2D_imbib}c and \ref{fig:2D_imbib}d in the conceptual model and $S_{r_{plane}}>0.8$ in the actual model, the effect of the projected areas at the top becomes more pronounced. This is because as the system reaches its air-expulsion value, the suction changes more dramatically (see, for instance, $S_r>0.95$ in Figure~\ref{fig:swcc}), which requires a dramatic change of menisci curvature. This is achieved by the rise of the liquid level at the surface, which reduces $A_{projected}$ and counteracts the effects of the pore filling at the bottom. In Figure \ref{fig:2D_imbib}c, the detachment of the gas cluster from grain 9 at the bottom increases the $A_{projected}$ by $0.05A$, but the rise of the liquid level at the surface more drastically decreases $A_{projected}$ by $0.07A$, resulting in an overall reduction of $\chi_{\Delta P}$. In Figure \ref{fig:2D_imbib}d, the removal of negative projected areas has canceled out with the decrease of projected areas at the top, resulting in $\chi_{\Delta P}$ remaining constant. Note that the trend at high saturations explained here is particular to the grain configuration in this study, since it has a large pore at the bottom that fills last, and can be more or less pronounced for other grain configurations.

At any given $S_{r_{plane}}$, $\chi_{\Delta P}$ is smaller during imbibition compared to drainage. We can explain this behavior by comparing the phase distributions in Figures~\ref{fig:2D_drain} and \ref{fig:2D_imbib}. Focusing on Figures ~\ref{fig:2D_drain}c and \ref{fig:2D_imbib}b, which have similar saturations, we identify two main consequences of the differences in phase distributions: 1) there are negative projected areas at the bottom during imbibition due to isolated gas clusters, and 2) most grains at the top have a smaller projected area during imbibition due to the liquid level at the surface being higher for the same $S_{r_{plane}}$. Both these      effects result in smaller $A_{projected}$ and, therefore, smaller $\chi_{\Delta P}$ during imbibition.

\noindent{4.4.2.2   $\chi_{\gamma_{lg}}$}

\begin{figure}
\centering
\includegraphics[]{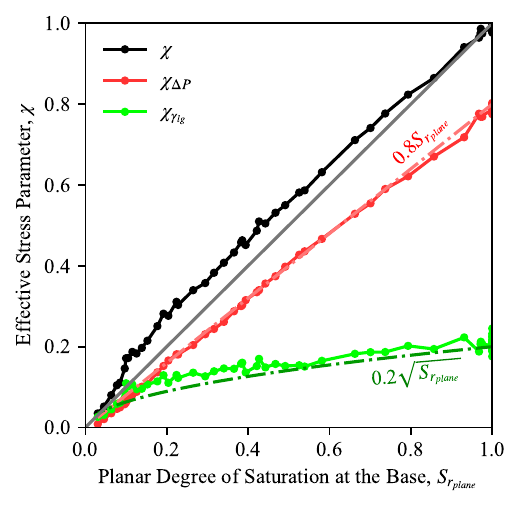}
\caption{Functional forms showing the general trends of $\chi_{\Delta P}$ and $\chi_{\gamma_{lg}}$ for the 3D granular soil column model during primary drainage.}
\label{fig:fitted_curve}
\end{figure}

\begin{figure}
\centering
\includegraphics[width=\textwidth]{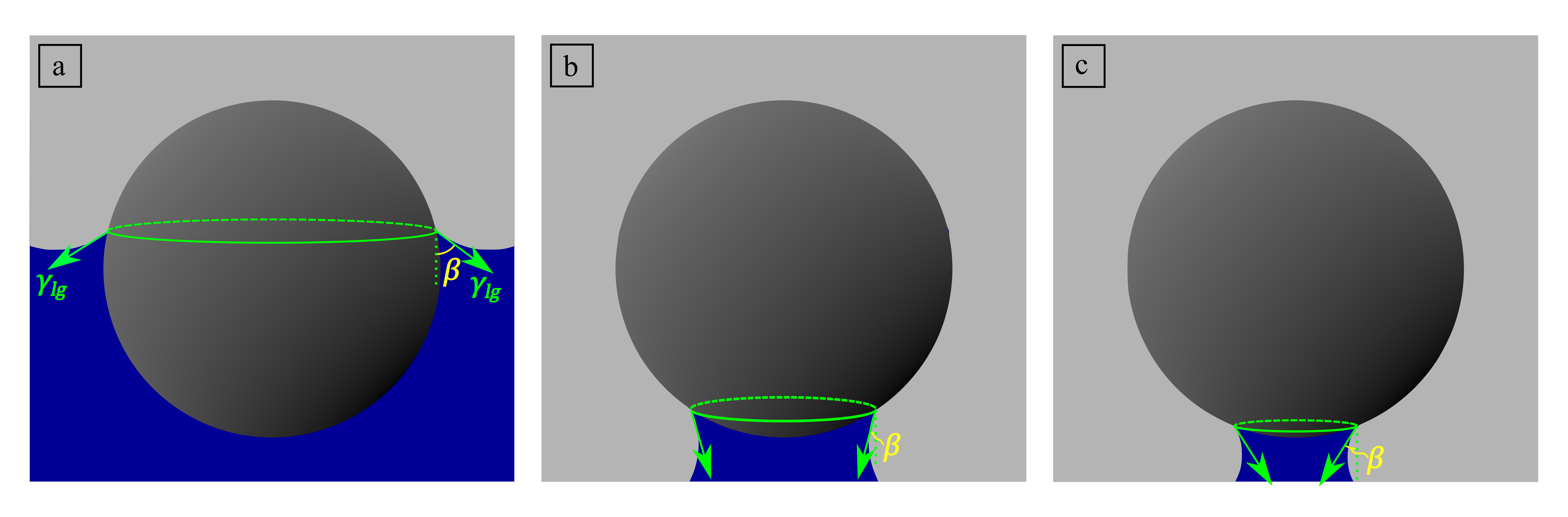}
\caption{A schematic diagram, of a single grain at the bottom of the soil column, for understanding the slope change in the $\chi_{\gamma_{lg}}$-${S_{r_{plane}}}$ curve.}
\label{fig:single_grain}
\end{figure}

As we saw previously in Figure~\ref{fig:chi}a, during primary drainage, $\chi_{\gamma_{lg}}$ decreases with $S_{r_{plane}}$ similar to $\chi_{\Delta P}$, but at a lower rate for $S_{r_{plane}}> 0.1$ and a higher rate for $S_{r_{plane}}< 0.1$. Based on Eq.~\ref{eq:chi_sum}, $\chi_{\gamma_{lg}}$ is affected by $F_{\gamma_{lg},z}$ and suction. For drainage, we find that the effect of $F_{\gamma_{lg},z}$ is more pronounced than the effect of suction. $F_{\gamma_{lg},z}$ itself is affected by the length of the contact line (e.g., $2\pi r$ in Figure~\ref{fig:schematic_forces}), as well as the orientation of the interfacial surface (e.g., $cos\beta$ in Figure~\ref{fig:schematic_forces}). The length of the contact line decreases with $S_{r_{plane}}$, for the same reason discussed previously for $A_{projected}$. However, since the contact line is one dimension lower than $A_{projected}$ (e.g., $2\pi r$ vs $\pi r^2$), its change is more closely proportional to $\sqrt{S_{r_{plane}}}$ rather than $S_{r_{plane}}$, as shown in Figure~\ref{fig:fitted_curve}. This is the main reason for the lower rate of $\chi_{\gamma_{lg}}$ compared to $\chi_{\Delta P}$. The second factor, which is the orientation of the interfacial surface, has different effects on $F_{\gamma_{lg},z}$ depending on the saturation level. To demonstrate this point, we utilize a schematic representation of a single grain at the bottom of the soil column in Figure~\ref{fig:single_grain}. At $S_{r_{plane}} = 1$, the grains at the bottom are either fully immersed in liquid or have a condition similar to Figure~\ref{fig:single_grain}a. As $S_{r_{plane}}$ decreases, these grains start transitioning from a condition similar to Figure~\ref{fig:single_grain}a to that of \ref{fig:single_grain}b. In this transition, the orientation of the interfacial surface changes such that $cos\beta$, which controls the magnitude of the vertical component, increases. This effect counteracts the decrease of the contact line length, and therefore, $F_{\gamma_{lg},z}$ and $\chi_{\gamma_{lg}}$ change at a low rate in this range. Once the base plane has entered the pendular regime, all grains at the bottom have a condition similar to \ref{fig:single_grain}b. At this point, further reduction in $S_{r_{plane}}$ requires the liquid bridges to shrink, similar to the transition from \ref{fig:single_grain}b to \ref{fig:single_grain}c. In this transition, $cos\beta$ and contact line length both decrease. Therefore, $F_{\gamma_{lg},z}$ and $\chi_{\gamma_{lg}}$ decrease at a higher rate in this range.

During secondary imbibition, Figure~\ref{fig:chi}b, $\chi_{\gamma_{lg}}$ is affected by the same factors discussed above; however, suction also plays a role in this case. Particularly at ${S_{r_{plane}}}>0.8$, where the system has reached its air-expulsion value and suction is dropping considerably, $\chi_{\gamma_{lg}}$ increases at a higher rate since $\chi_{\gamma_{lg}}$ is inversely proportional to the suction magnitude.

At any given ${S_{r_{plane}}}$, $\chi_{\gamma_{lg}}$ is larger during imbibition compared to drainage. A large part of this is due to the suction being lower during imbibition (i.e., SWCC hysteresis). However, $F_{\gamma_{lg},z}$ also plays a role by being higher during imbibition, as seen in Figure~\ref{fig:forces}c2. Let us refer back to Figures~\ref{fig:2D_drain}c and ~\ref{fig:2D_imbib}b, this time focusing on $\boldsymbol{f_{\gamma_{lg}}}$ vectors for the grains at the bottom. We see that the particular phase distribution during imbibition makes the orientation of the interfacial surfaces such that $\boldsymbol{f_{\gamma_{lg}}}$ pulls the grain towards the gas phase, unlike during primary drainage where $\boldsymbol{f_{\gamma_{lg}}}$ pushes the grain away from the gas phase. The addition downward $\boldsymbol{f_{\gamma_{lg}}}$ during imbibition results in a higher $F_{\gamma_{lg},z}$ and contributes to the higher $\chi_{\gamma_{lg}}$.

\subsubsection{Comparison of the Numerically Measured and the Theoretically Calculated $\chi$}
\label{comparison}

\begin{figure}
\centering
\includegraphics[]{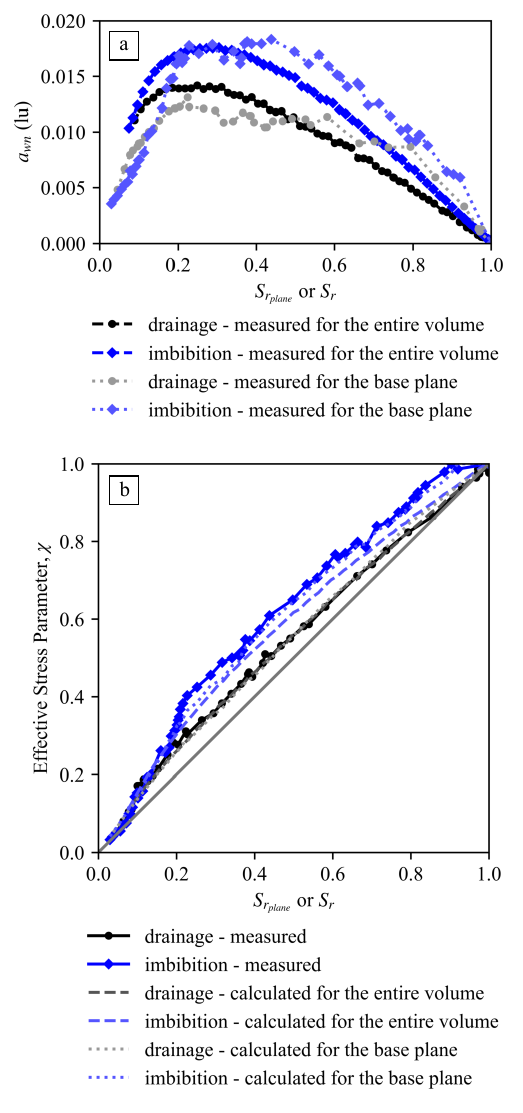}
\caption{a) Measured specific interfacial area, $a_{wn}$ and b) comparison of numerically measured and theoretically calculated $\chi$, for the 3D granular soil column model.}
\label{fig:comparison}
\end{figure}

Nikooee at el. \citeyear{nikooee_effective_2013} derive $\chi$ as 
\begin{equation}
\chi=S_r + \frac{k_{wn}a_{wn}}{\Delta P}
\label{eq:chi_theory}
\end{equation}
where $k_{wn}$ is a material coefficient and $a_{wn}$ is the wetting-nonwetting (liquid-gas) interfacial area per volume. Likos \citeyear{likos_effective_2014} suggests $k_{wn} = \gamma$, where $\gamma$ is the surface tension we have been referring to as $\gamma_{lg}$. The particular difficulty when using Eq.~\ref{eq:chi_theory} is that $a_{wn}$ cannot be easily measured or calculated. However, we are able to measure $a_{wn}$ at each hydraulic step based on the multiphase LBM results. $a_{wn}$ for primary drainage and secondary imbibition is presented in Figure~\ref{fig:comparison}a. The reason for the higher $a_{wn}$ during secondary imbibition goes back to the difference in phase distributions discussed in \citeN{hosseini_investigating_2024}. If we substitute the measured $a_{wn}$ and $S_r$ into Eq.~\ref{eq:chi_theory}, with $k_{wn} = \gamma$, we find the $\chi$ corresponding to the average effective stress for the volume, demonstrated with dashed lines in Figure~\ref{fig:comparison}b. We see that the calculated solution matches the measured results perfectly for primary drainage but not for secondary imbibition. As mentioned previously, our method of measuring $\chi$ corresponds to the effective stress on the base plane, not the average over the volume. If instead of $a_{wn}$ for the entire soil column, we only measure the length of the interfacial line on the base plane, for example the green line in Figure~\ref{fig:bottom_plane}, and divide it by the cross-sectional area rather than the volume, we get the curves shown with dotted lines in Figuree~\ref{fig:comparison}a. If we now use these values in Eq.~\ref{eq:chi_theory} alongside ${S_{r_{plane}}}$, we get the curves shown with dotted lines in Figure~\ref{fig:comparison}b. These curves match our numerical measurements very well, showing that the hysteresis in $\chi$ we have measured is consistent with the theory.

\section{Conclusion}

We gain insight into the effective stress parameter, $\chi$, in unsaturated soils by investigating the influence of suction and surface tension forces at the grain scale. We develop an algorithm to calculate these forces using multiphase lattice Boltzmann method (LBM) simulations. We simulate primary drainage and secondary imbibition curves for a synthetic 3D granular soil column and measure $\chi$ along the way. We see that $\chi=S_r$ only applies to $S_r=1$ and $S_r=0$. At all other saturation levels, $\chi$ is greater than $S_r$. The maximum deviation of $\chi$ from $S_r$ happens at the transition point to/from the pendular regime, where the multitude of large liquid bridges induce large surface tension forces. We also see that the $\chi$-$S_r$ relationship is hysteretic outside the pendular regime, where $\chi$ is larger during imbibition compared to drainage. Investigating the different components that contribute to $\chi$, we find that: 1) the contribution of suction forces, $\chi_{\Delta P}$, is lower during imbibition compared to drainage, 2) the contribution of surface tension forces, $\chi_{\gamma_{lg}}$, is higher during imbibition compared to drainage, and 3) among the two components, the hysteresis effect of $\chi_{\gamma_{lg}}$ is more pronounced than $\chi_{\Delta P}$, resulting in the overall $\chi$ to be higher during imbibition. To justify these findings, we take advantage of conceptual models, which are schematic representations of the granular soil. We show that the isolated gas clusters during imbibition reduce $\chi_{\Delta P}$, while increasing $\chi_{\gamma_{lg}}$. Finally, we compare our results with a theoretical solution by substituting the specific interfacial area, $a_{wn}$, measured in our simulations into the proposed equation by \citeN{nikooee_effective_2013}. We find a good match between our micromechanical evaluation of $\chi$ and the theoretical solution.

\section*{Author Statements}

\subsection*{Competing Interests}
\noindent The authors declare there are no competing interests.

\subsection*{Author Contributions}
\noindent Conceptualization: RH, KK\newline
Formal analysis: RH\newline
Funding acquisition: KK\newline
Investigation: RH\newline
Methodology: RH\newline
Software: RH\newline
Supervision: KK\newline
Validation: RH\newline
Visualization: RH\newline
Writing – original draft: RH\newline
Writing – review and editing: KK

\subsection*{Funding}
\noindent The authors declare no specific funding for this work.

\subsection*{Data Availability}
\noindent The data is available upon request from the corresponding author.

\pagebreak

\bibliography{stress_paper}

\pagebreak
%
\appendix
\renewcommand\thefigure{\thesection\arabic{figure}}    
\renewcommand\theequation{\thesection\arabic{equation}}  
\setcounter{figure}{0} 
\setcounter{equation}{0} 

\section{Contact angle measurement}
\label{app:contact angle}

\begin{figure}[h]
\centering
\includegraphics[]{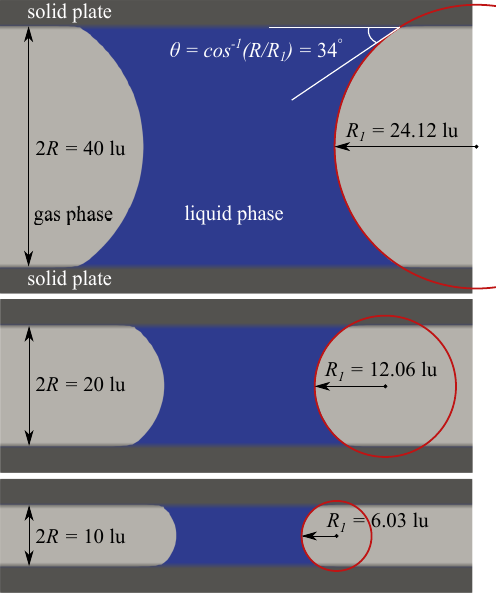}
\caption{Contact angle measurement using 2D simulations of a liquid bridge between two solid plates. Three different plate spacings are used, all of which result in a contact angle of $34^{\circ}$ for the LBM model with $\rho_s=0.25$.}
\label{fig:contact_angle}
\end{figure}

We use the simulation of a liquid bridge between two solid plates to measure the contact angle and surface tension for a given solid density, $\rho_s$, as discussed in \citeN{hosseini_investigating_2024}. We fit a circle to the surface of the liquid bridge and measure the radius of the fitted circle, $R_1$. We then find the contact angle, $\theta$, using $\theta = cos^{-1}(R/R_1)$, where $R$ is half the distance between the two plates. Figure~\ref{fig:contact_angle} shows this procedure for three different plate spacings.

\section{Validation of the fluid force measurement algorithm}
\label{app:validation}
\setcounter{figure}{0} 
\setcounter{equation}{0} 

To validate the algorithm presented in Section~\ref{algorithm}, we simulate a liquid bridge between two spherical grains and compare the forces between numerical simulations and analytical calculations.

\subsection{Analytical Solution}
\begin{figure}[h]
\centering
\includegraphics[]{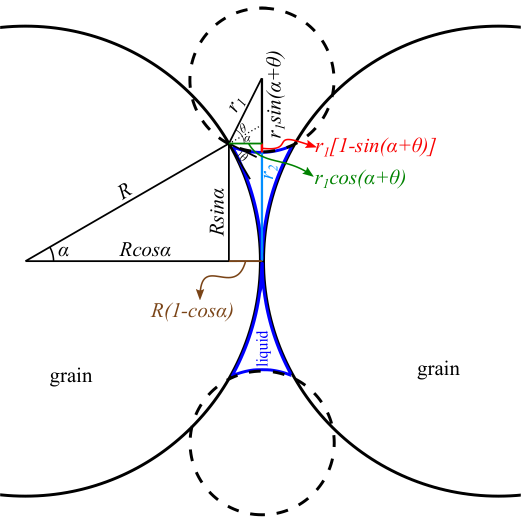}
\caption{Graphical procedure for finding the radii of curvature of a liquid bridge between two grains as a function of the contact angle and the filling angle.}
\label{fig:analytical}
\end{figure}

Based on Figure~\ref{fig:analytical}, which shows a 2D cross-section of a liquid bridge between two spherical grains, we can derive the radii of curvature, $r_1$ and $r_2$, as a function of the contact angle, $\theta$, and the filling angle, $\alpha$:
     
\begin{equation}
R(1-cos\alpha)=r_1cos(\alpha+\theta) \rightarrow r_1=R(1-cos\alpha)/cos(\alpha+\theta)
\label{eq:r1}
\end{equation}

\begin{equation}
R sin\alpha=r_2+r_1[1-sin(\alpha+\theta)] \rightarrow r_2=R sin\alpha-r_1[1-sin(\alpha+\theta)]
\label{eq:r2}
\end{equation}
     
The suction force and the surface tension force applied to a single grain can be derived as
\begin{equation}
F_{\Delta P}=(P_g - P_l)\times \pi(R sin\alpha)^2
\label{eq:f_suction1}
\end{equation}

\begin{equation}
F_{\gamma_{lg}}=\gamma_{lg}\times2\pi R sin\alpha \times sin(\alpha + \theta)
\label{eq:f_tension}
\end{equation}
which are oriented horizontally since all other components cancel out. In addition, based on the Young-Laplace equation, we know that
\begin{equation}
\Delta P=P_g - P_l=\gamma_{lg}(\frac{1}{r_1}-\frac{1}{r_2}),
\label{eq:dp}
\end{equation}
where $r_1$ and $r_2$ are absolute values and therefore a negative sign has been incorporated into the equation to account for the proper sign of $r_2$. Substituting Eq.~\ref{eq:dp} in Eq.~\ref{eq:f_suction1}, we find
\begin{equation}
F_{\Delta P}=\gamma_{lg}(\frac{1}{r_1}-\frac{1}{r_2})\times \pi(R sin\alpha)^2
\label{eq:f_suction2}
\end{equation}
We use Eq.~\ref{eq:f_suction2} and Eq.~\ref{eq:f_tension}, in the following section, to calculate the analytical solutions for the suction and surface tension forces.

\subsection{Numerical Simulations}

\begin{figure}[h]
\centering
\includegraphics[]{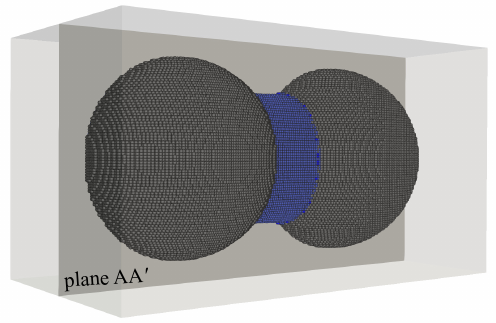}
\caption{Multiphase LBM simulation of a liquid bridge between two spherical grains, with radii of 40 lu, for validation of the fluid force measurement algorithm.}
\label{fig:two_grains}
\end{figure}

\begin{figure}[]
\centering
\includegraphics[width=\textwidth]{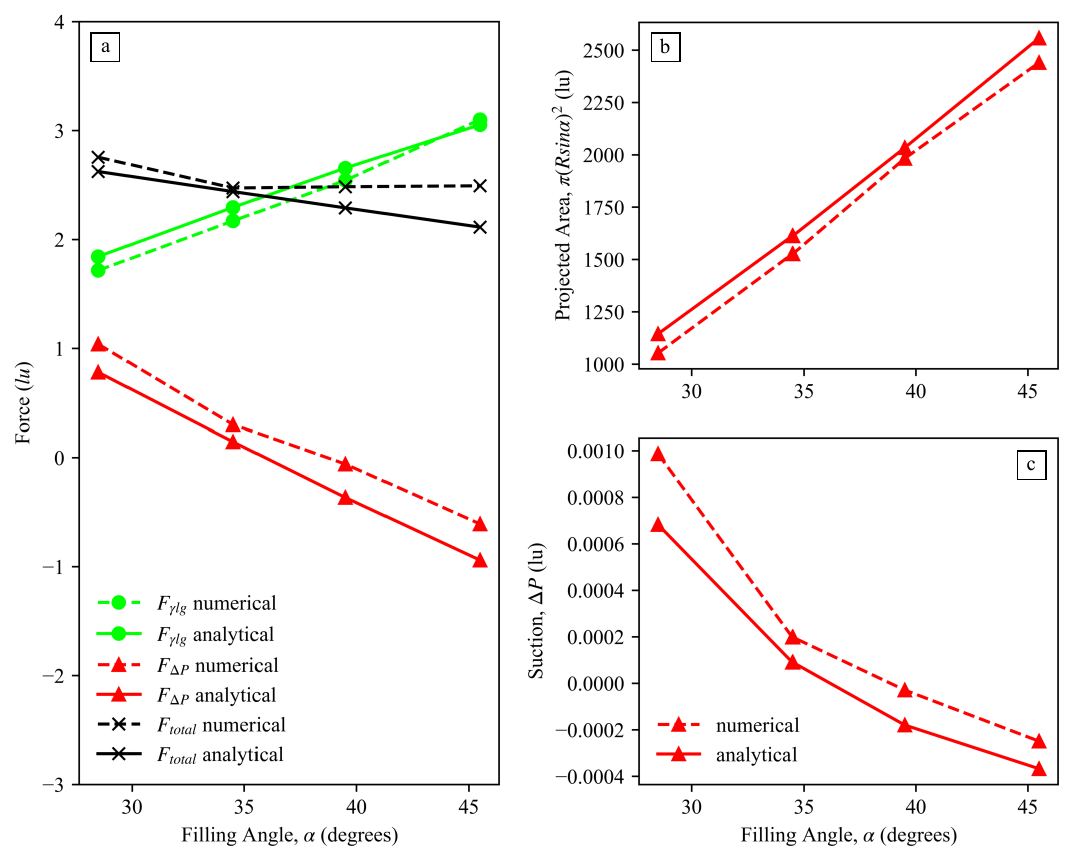}
\caption{Comparison of the numerical measured and analytically calculated results for the liquid bridge between two spherical grains model.}
\label{fig:num_results}
\end{figure}

\begin{figure}[]
\centering
\includegraphics[]{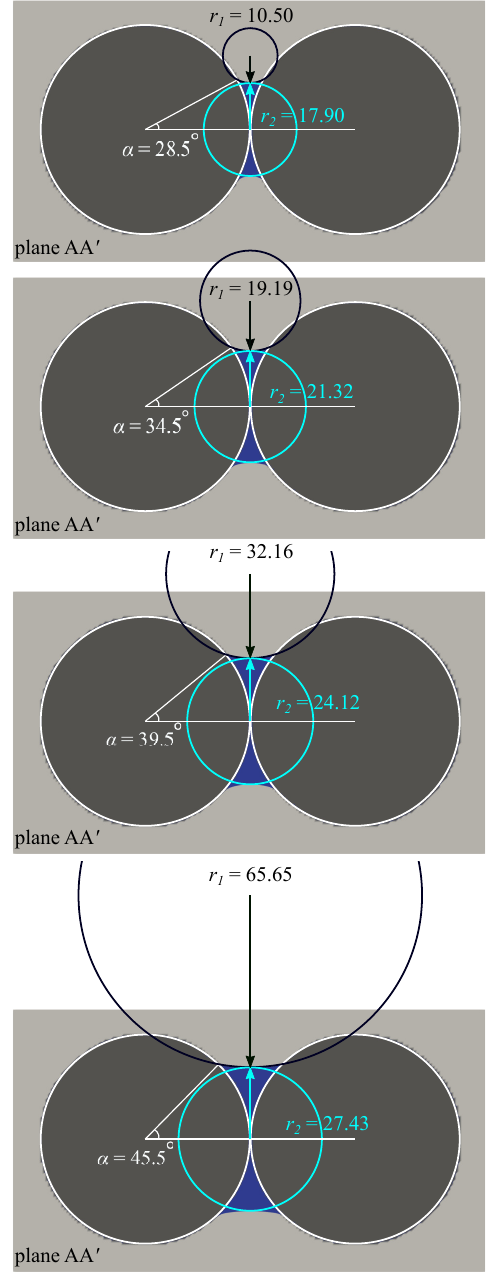}
\caption{Finding the filling angle, $\alpha$, and radii of curvature, $r_1$ and $r_2$, for liquid bridges between two grains simulated using multiphase LBM.}
\label{fig:filling_angle}
\end{figure}

We simulate two grains with radii of 40 lu with four different bridge sizes between them, the largest of which is shown in Figure~\ref{fig:two_grains}. For each simulation, we measure the suction and surface tension forces numerically based on the procedure outlined in Section~\ref{algorithm}. These results are shown with dashed lines in Figure~\ref{fig:num_results}. To find the analytical solution, for each simulation, we plot the phase distributions on a cross-section passing through the center of the grains and graphically find the filling angle, as shown in Figure~\ref{fig:filling_angle}. By knowing the grain radius, the contact angle and the filling angle, we use Eq.~\ref{eq:r1} and Eq.~\ref{eq:r2} to find the radii of curvature, and we verify them graphically, as seen in Figure~\ref{fig:filling_angle}. We then use Eq.~\ref{eq:f_suction2} and Eq.~\ref{eq:f_tension} to find the analytical solution for suction force and surface tension force, respectively. These results are shown with solid lines in Figure~\ref{fig:num_results}. Comparing the numerical solution to the analytical solution in Figure~\ref{fig:num_results}a, we see that in our numerical simulations, the surface tension force is underestimated, on average by 5\%, while the suction force is overestimated, on average by 60\%. 

To understand the underestimation of the surface tension force, we need to investigate the algorithm used for contact line detection. While in reality, the contact line can be anywhere in the continuous space, in the numerical simulation, the space is discretized into lattice nodes, thus restricting the location of the contact line to these nodes. Therefore, when it comes to identifying the location of the contact line, we have two options: choosing liquid nodes with neighboring gas nodes, or choosing gas nodes with neighboring liquid nodes; while in reality, the location is somewhere between these liquid and gas nodes. As explained in Section~\ref{surface tension force}, we choose the first option. Therefore, the contact line we find is inside the liquid domain, making it smaller than the actual contact line. Alternatively, if we identify the contact line using gas nodes, the contact line will be slightly larger than the true one. Of course, for better estimation, we can use both methods and take the average of the results; however, that would significantly increase the computational cost of larger simulations in subsequent sections. Therefore, we accept the -5\% error for this study.

To understand the overestimation of the suction force, we need to look into the components of the suction force: suction, $\Delta P$, and the projected area it acts on, $\pi(R sin\alpha)^2$. Plotting the numerically measured and analytically calculated values for each component individually in Figures~\ref{fig:num_results}b and \ref{fig:num_results}c reveals that the error in the suction force measurement is mainly due to an error in the suction measurements. In fact, our algorithm for measuring the surface areas is producing relatively accurate results with an average error of -5\%, with the same reason for the underestimation as explained for the contact line detection. Our investigation shows that the cause of the overestimation of the suction (70\% on average) is simply the grid resolution. Suction is the difference between gas and liquid pressures. While the variation in gas pressure is negligible in these simulations, the liquid pressure has large variations from negative to positive values based on the equation of state (see Figure 3 in hysteresis paper). Due to the diffuse liquid-gas interface in multiphase LBM (i.e., gradual change of fluid density from gas density to liquid density), the liquid has low pressures near the boundary, and the pressure gradually builds up toward the coexistence pressure (i.e. pressure at which liquid and gas phases coexist) when moving away from the boundary. Therefore, for an accurate estimate of the liquid pressure, there should be a sufficient number of lattice nodes inside the liquid zone to allow the buildup of liquid pressure. If there are fewer lattice nodes than necessary, the liquid pressure may not build up properly before encountering another boundary that decreases the pressure. Our results (not shown herein for brevity) indicate that the measured suction converges towards the calculated suction with an increase in the grid resolution (i.e. grains of radius 10 lu vs 20 lu vs 40 lu). However, currently, due to our computational limitations, we cannot increase the resolution of the large simulations further and accept the overestimation. Although the suction forces are overestimated, multiphase LBM captures the trends correctly, which is the primary focus of the paper.

\end{document}